# An Open-Sourced, Community-Driven Volumetric Additive Manufacturing Printer and Post-Processor

International Symposium on Academic Makerspaces

Taylor Waddell[1], Erik Broude[1], Tristan Bourgade[1], Natalia Fabiana De La Torre[1], Erfan Kohyarnejadfard[1], Tavleen Kaur[1],
Scarlett Hao[1], Dylan Motley[1], Daniel Oslund[1], Evan Percival[1], Rajdeep Summan[1], Connor Vidmar[1], Hayden Taylor[1]

[1] Department of Mechanical Engineering, University of California, Berkeley, Berkeley, CA, 94720, USA

**Abstract**

Volumetric Additive Manufacturing (VAM) represents a recent advancement in additive manufacturing (AM), offering several advantages over conventional methods such as Fused Filament Fabrication (FFF) and Stereolithography (SLA). These advantages include layer-less fabrication, reduced reliance on wasteful support structures, and significantly faster print speeds. Additionally, VAM enables the use of higher-viscosity precursor materials, as the printed object remains stationary relative to the photopolymer volume during fabrication.

A specific type of VAM, known as Computed Axial Lithography (CAL) [1], delivers a spatially modulated light dose into a photopolymer resin using tomographic reconstruction (Figure 1). CAL offers significant potential for accessibility due to its mechanical simplicity; in its most basic form, the system requires only a visible-light projector, a vial of photopolymer, and a rotation mechanism. The underlying tomographic reconstruction algorithms necessary for CAL have been implemented in the open-source software package VAMToolbox [2], which enables the generation of projection images required to deliver the appropriate dose distribution for part formation.

OpenCAL is a low-cost CAL-based printing and post-processing platform developed using commercial off-the-shelf (COTS) components and standard rapid prototyping tools. The system incorporates recent advancements in VAM, including Optical Scattering Tomography (OST) [3], and is designed with modularity to support future technological integration. The post-processing systems developed include standardized procedures for solvent-based removal of uncured resin and a centrifugal cleaning module for enhanced material recovery.

To support the continued advancement of this technology, multiple community engagement pathways have been established to foster the collaborative development of the OpenCAL ecosystem. A primary deployment environment for OpenCAL is academic makerspaces, where the system is specifically designed to leverage the rapid prototyping tools typically available in these facilities. OpenCAL not only introduces advanced volumetric additive manufacturing capabilities to makerspaces but also serves as a platform for hands-on education in emerging manufacturing technologies, photopolymer science, and computational imaging. By lowering the barriers to entry for volumetric printing research, OpenCAL enables a broader range of students—from undergraduate to graduate levels—to engage directly in experimental research, contribute to the refinement of open-source CAL technologies, and participate in interdisciplinary projects spanning materials science, mechanical engineering, and computer science. In doing so, OpenCAL has the potential to significantly expand the technical capabilities of academic makerspaces, transforming them into hubs for next-generation manufacturing innovation and research-driven learning.

## 1. Introduction & Related Work

Community-driven additive manufacturing initiatives have historically served as catalysts for research and innovation within the field [4]. With growing interest from hobbyists and educators in obtaining access to emerging manufacturing technologies, community-driven projects possess substantial potential for impact and widespread adoption. In particular, for early-stage emerging consumer technologies, open-source and community-led efforts are often critical to achieving long-term success, as exemplified by the Arduino or Reprap projects [5], [6]. Computed Axial Lithography (CAL) offers the opportunity to establish a new frontier in additive manufacturing; however, its adoption has traditionally been limited to communities with access to substantial funding and advanced scientific infrastructure [1]. Although several open-source CAL demonstrations have been undertaken, they have struggled to gain widespread traction due to challenges related to accessible materials, lack of organized community engagement, and hardware scalability limitations [7].

OpenCAL seeks to address these barriers by developing a fully accessible, low-cost CAL platform that can be utilized and expanded by a broad range of users. A primary deployment target for OpenCAL is academic makerspaces, which offer a uniquely valuable environment for the incubation of emerging technologies. While makerspaces provide students with access to fabrication tools and prototyping equipment, they often struggle to deliver deeply rewarding research experiences in highly technical fields due to resource limitations, lack of specialized expertise, and difficulty maintaining cutting-edge equipment. OpenCAL offers a means to overcome these challenges by introducing a research-grade volumetric manufacturing capability into makerspaces, enabling students to engage directly with advanced fabrication processes and contribute to open-source development efforts.

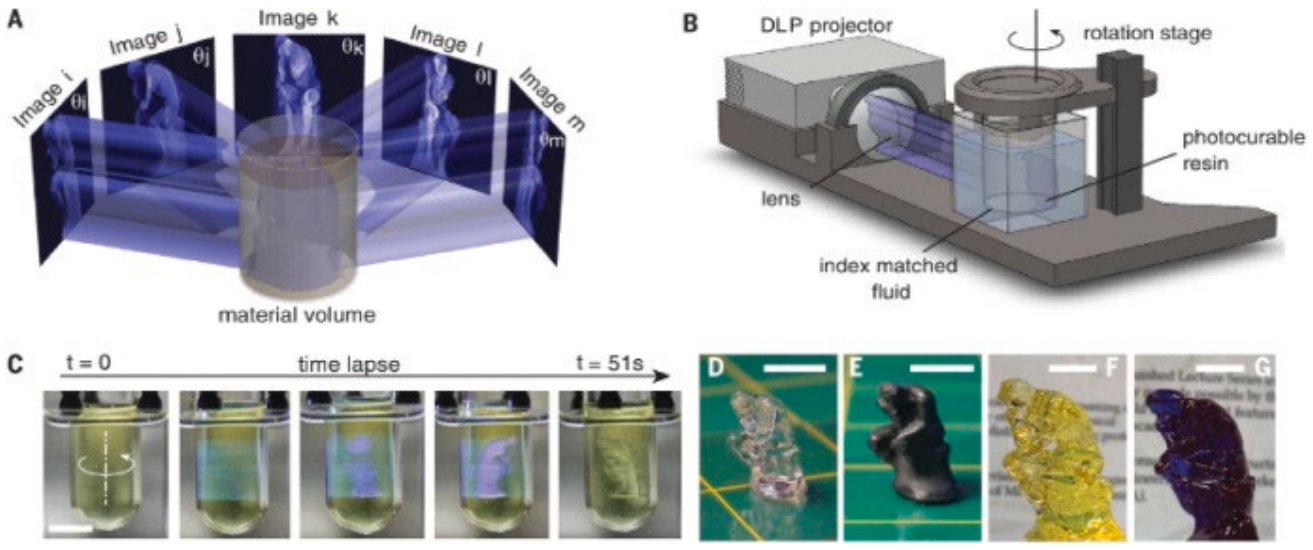

*Figure 1: CAL volumetric fabrication. (A) Underlying concept: patterned illumination from many directions delivers a computed 3D exposure dose to a photoresponsive material. (B) Schematic of CAL system used in this work. (C) Sequential view of the build volume during a CAL print. A 3D geometry is formed in the material in less than a minute. (D) The 3D part shown in (C) after rinsing away uncured material. (E) The part from (D) painted for clarity. (F) A larger (40 mm-tall) version of the same geometry. (G) Opaque version of the geometry in (F), using crystal violet dye in the resin. Scale bars: 10 mm. Reproduced from Kelly et al. [1] with the permission of the American Association for the Advancement of Science. (For interpretation of the references to color in this figure legend, the reader is referred to the Web version of this article.)*

Moreover, academic makerspaces are frequently situated near research laboratories, providing a unique opportunity for integration between hands-on fabrication environments and formal scientific research. OpenCAL offers a platform to establish collaborative partnerships between makerspaces and research groups, specifically targeting the advancement of VAM. Many research laboratories remain unfamiliar with CAL-based methodologies or are not equipped to support volumetric fabrication. By deploying OpenCAL within makerspaces, students can facilitate exploratory research and contribute experimental data critical to advancing the field. This fosters a symbiotic relationship in which makerspaces gain increased technical depth and research relevance, while laboratories benefit from an engaged and skilled talent pipeline capable of accelerating CAL technology development. In this framework, OpenCAL has the potential to assist academic makerspaces to become stronger innovation hubs that actively bridge education, research, and the community-driven evolution of advanced manufacturing technologies.

## 2. Methods

### 2.1 Initial Development

The development of the OpenCAL platform followed a structured product development process, encompassing phases of customer discovery, design exploration, prototyping, testing, and deployment. Customer discovery engaged a range of stakeholders, including individuals both familiar and unfamiliar with VAM technologies (Figure 2). A total of 22 interviews were conducted, resulting in the identification of key product requirements and user concerns. The most prominent concern expressed was the handling and management of liquid chemicals, particularly printing resins and post-washing solvents. Additional critical requirements identified included a target system cost below $1000 USD, an assembly time of less than one day, and operational readiness to begin printing within two hours of assembly. Internal technical specifications were derived including a maximum resin container ("vial") diameter of 4 inches, permissible rotational runout of 100 microns, and target vial rotation speeds between 2–10 RPM. The diverse participant pool informed a comprehensive, user-centered system design. Additionally, the diversity of interviewees ensured integration of broad yet specific user needs. For example, the requirement for modularity was incorporated to address the needs of academic researchers, while the fabrication process was constrained to the use of general-purpose prototyping tools—such as 3D printers, laser cutters, and soldering equipment—reflecting the typical capabilities of academic makerspaces.

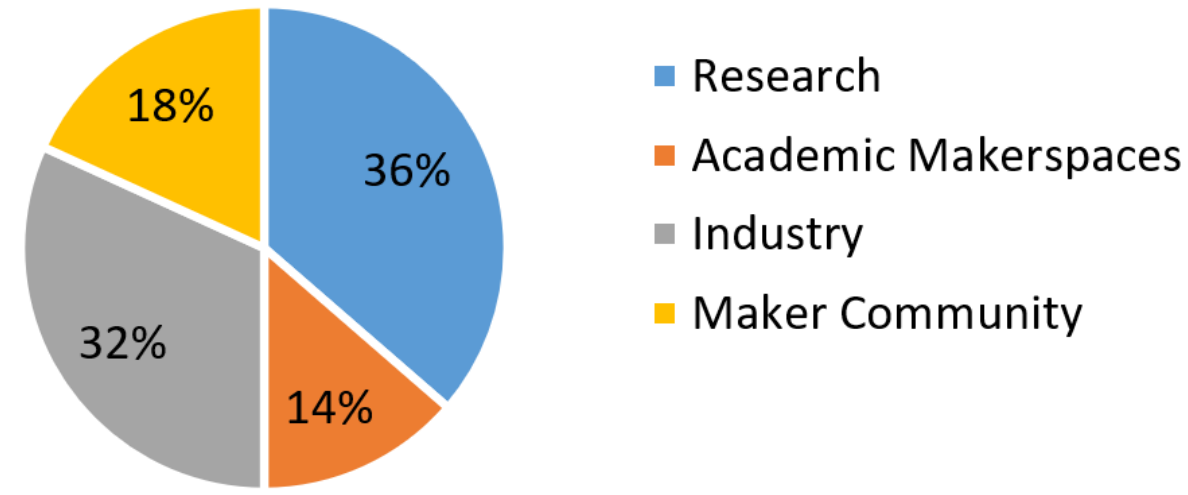

*Figure 2: Background of interviewees during customer discovery.*

Design exploration began with designers actively participating in a complete CAL printing workflow. This process included loading precursor material into the printing container, printing, removing the printed part, cleaning the part to remove uncured resin, and performing post-curing. Direct participation in the full workflow allowed students to identify key pain-points of the technology that need improvement and greater accessibility for wider adoption.

Based on the initial evaluation, three primary development areas were identified. The first focused on the optical system, specifically the adaptation of a commercially available DLP projector for CAL-based printing.

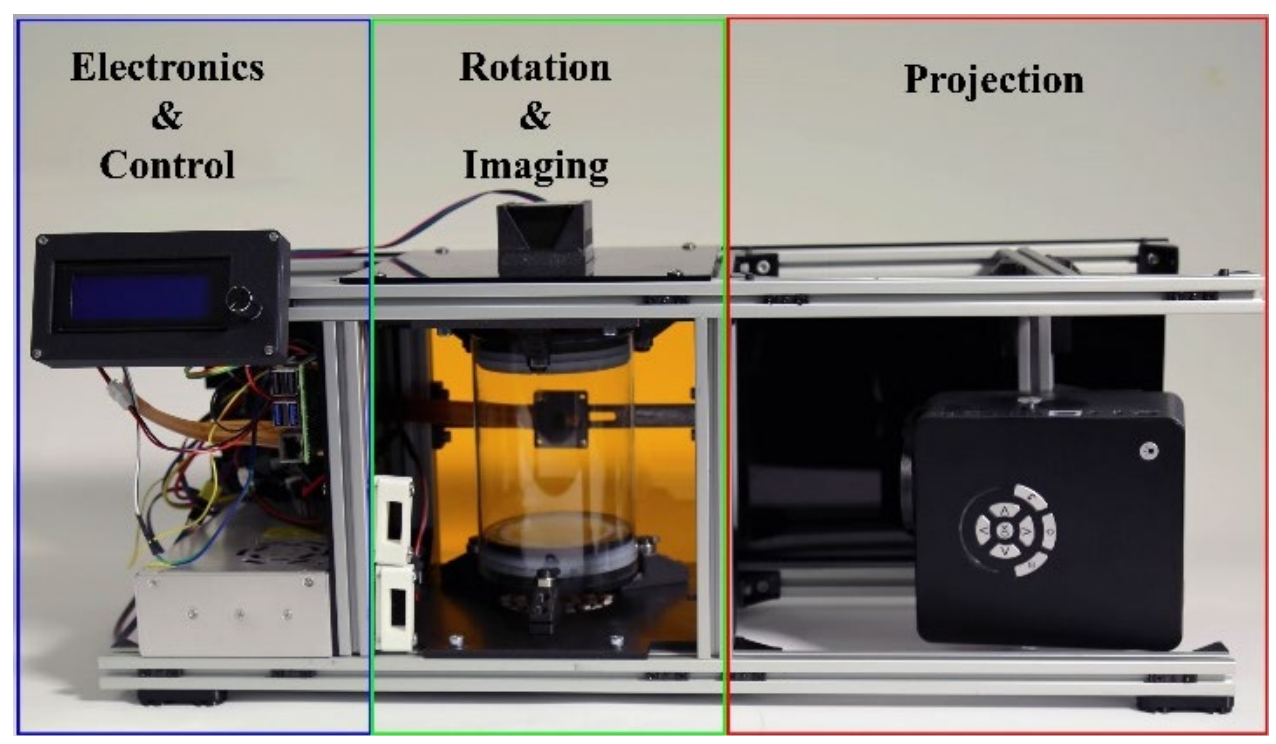

*Figure 3: DLP-based OpenCAL system with core components.*

The second major focus area addressed the development of a resin handling and post-processing system. This included filling the printing vials, removing residual fluid, cleaning printed parts, and performing post-curing. Two post-processing approaches were investigated: a solvent-based method, where excess resin is dissolved using chemical baths, and a centrifugal method, where centrifugal forces are applied to the vial and part to remove excess resin, leaving a thin film of remaining material.

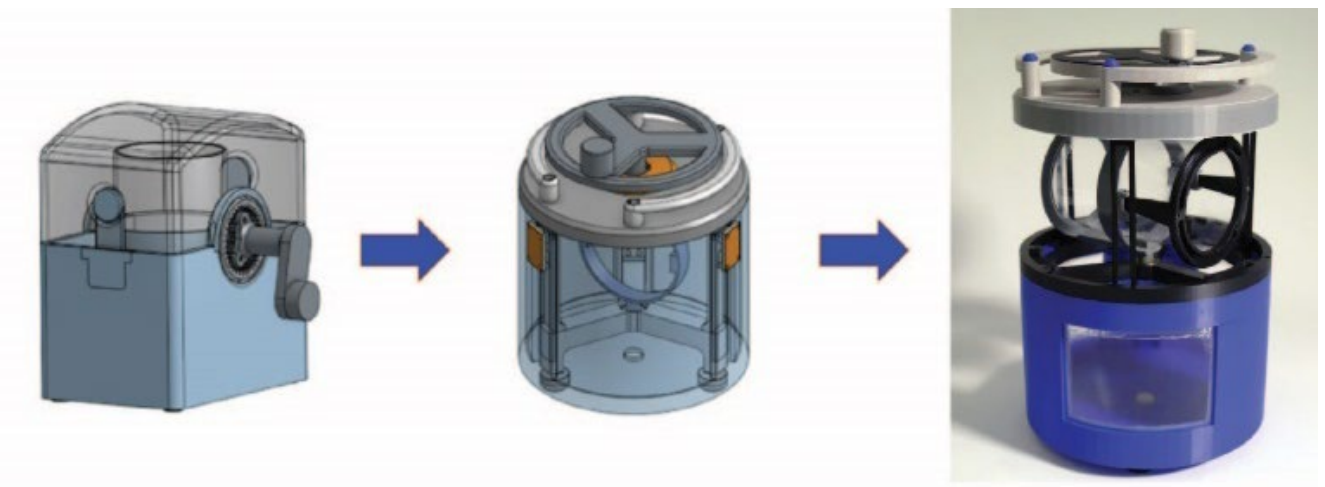

*Figure 4: Evolution of the centrifugal post-processing system.*

The third focus area involved identifying an accessible photopolymer resin suitable for OpenCAL. No COTS material had previously been demonstrated to function effectively with VAM, necessitating the development of custom material solutions [8]. This research effort systematically evaluated existing COTS photopolymers to determine their compatibility with the VAM process and identify viable candidates for OpenCAL applications.

### 2.2 Core System Development

The core system contains the vial, the rotational element, the main structure, the camera system, and the electrical system. The vial system was developed to have a transparent bottom for metrology illumination [3]. The top lid would be the adapter for rotation, and both lids could be removed for cleaning purposes. The vial system was designed to be adaptable for different vials, demonstrated with an adapter for a 30 mm-diameter vial system.

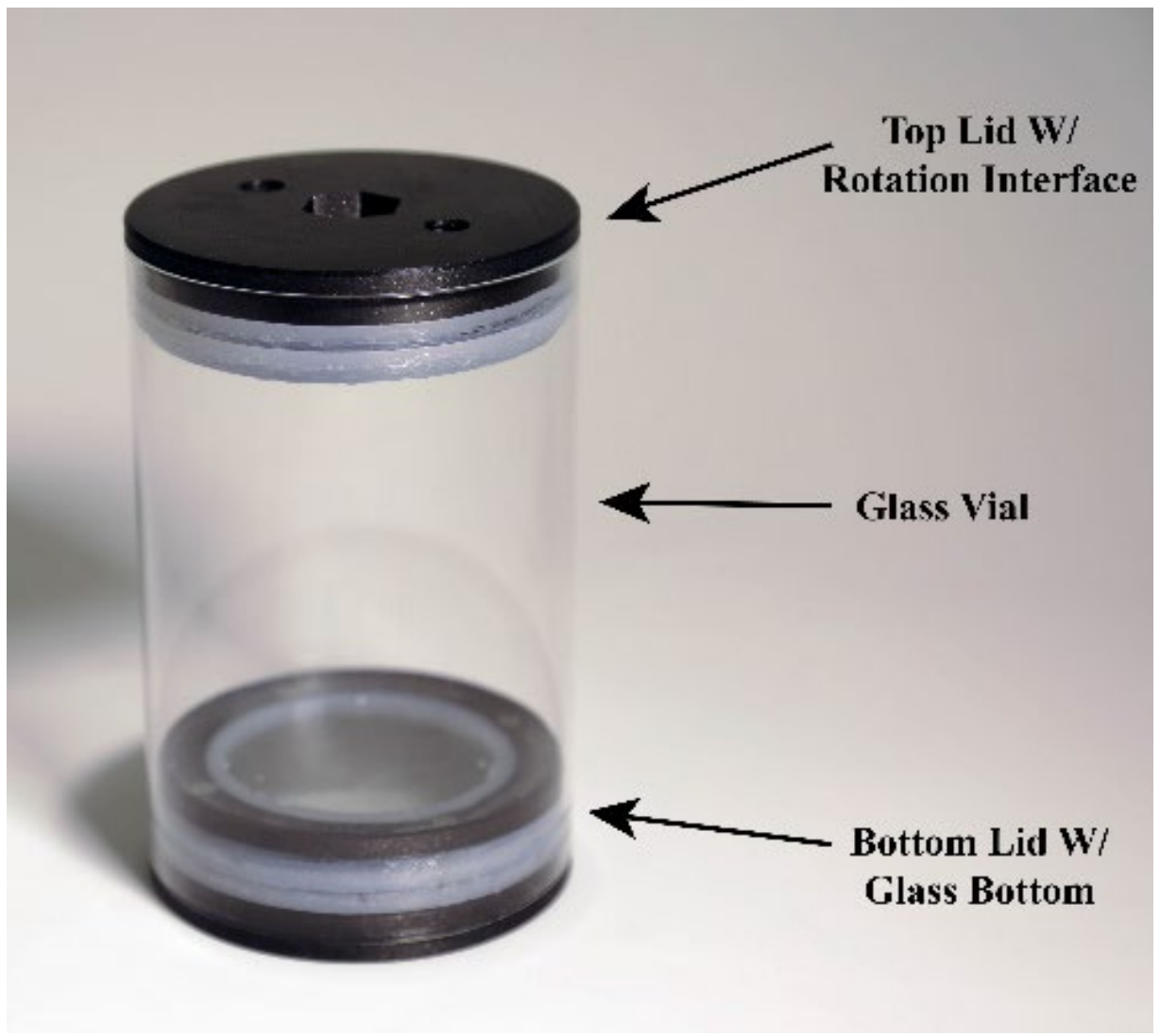

*Figure 5: A 4 inch-diameter, 6 inch-tall standard OpenCAL vial.*

The structure was designed mainly out of 80/20 extrusion (80/20 corporation, Columbia City, IN), as well as 3D-printed parts and laser-cut sheets as seen in Figure 3.

The electrical system utilizes only easily accessible COTS parts, and it is primarily based on a Raspberry Pi 5 microprocessor [9]. The camera and data collection system could utilize a USB camera or a Raspberry Pi ribbon-cable camera.

### 2.3 Optical Development

Together, the structural, electrical, and control subsystems provided a robust foundation for integrating the optical subsystem. The AAXA P6 Ultimate projector [10] was selected for its high LED-based optical output of 1100 lumens, relatively low cost of $369 USD compared to similar projectors, and anticipated long-term manufacturer support. Two immediate modifications were required for the projector's integration. First, keystone correction had to be addressed. Due to the projector's 100% offset—meaning the bottom edge of the projected image aligns with the center of the projector lens—it had to be physically angled relative to the vial's vertical axis to align the optical path through the center of the vial. This angular positioning introduced geometric distortion in the projected image. To correct this distortion, the projector's internal computational keystoning was utilized, and the correction settings were calculated and applied accordingly.

Second, the throw ratio of the projector needed to be increased to reduce the divergence of the projected rays. Minimizing beam divergence is critical for CAL, as the projection algorithms implemented in VAMToolbox assume a collimated light source. To address this need, a 250 mm focal length lens was added to the end of the projector's optical path, successfully reducing divergence and shortening the effective focal length of the projected image. Optical intensity measurements across the projector's emission spectrum were conducted, with a peak intensity of 9 mW/cm² observed at 450 nm. Accordingly, 450 nm was selected as the new operational printing wavelength for the OpenCAL system.

### 2.4 Post-Processing Development

The primary method for post-processing CAL parts involves submerging them in a solvent, typically isopropyl alcohol (IPA), followed by agitation through mixing and subsequent drying. However, this method often erodes fine features on the parts and generates substantial quantities of hazardous waste. To address these challenges, ultrasonic cleaning was investigated as an alternative post-processing approach. In this method, mechanical forces generated by sonication are used to remove excess resin from the printed parts.

Initial tests were conducted in a pure water solution, where parts were suspended on a mesh and subjected to ultrasonic agitation for 30 minutes. For convex geometries, this method effectively removed excess resin without compromising fine surface features. However, for concave geometries, residual resin remained trapped within recessed features. To improve resin removal in these areas, ethanol was added to the water solution at concentrations of up to 25% by volume. Ethanol was selected for its ability to gently dissolve hydrophobic resin materials, in contrast to more aggressive solvents such as acetone and IPA, and for its widespread availability. The modified ultrasonic cleaning process demonstrated promising results, successfully cleaning parts with a range of geometrical complexities, as illustrated in Figure 6. While these experiments were initially conducted with a laboratory-grade ultrasonic cleaner, they were recreated with success using a cheap COTS ultrasonic jewelry cleaner.

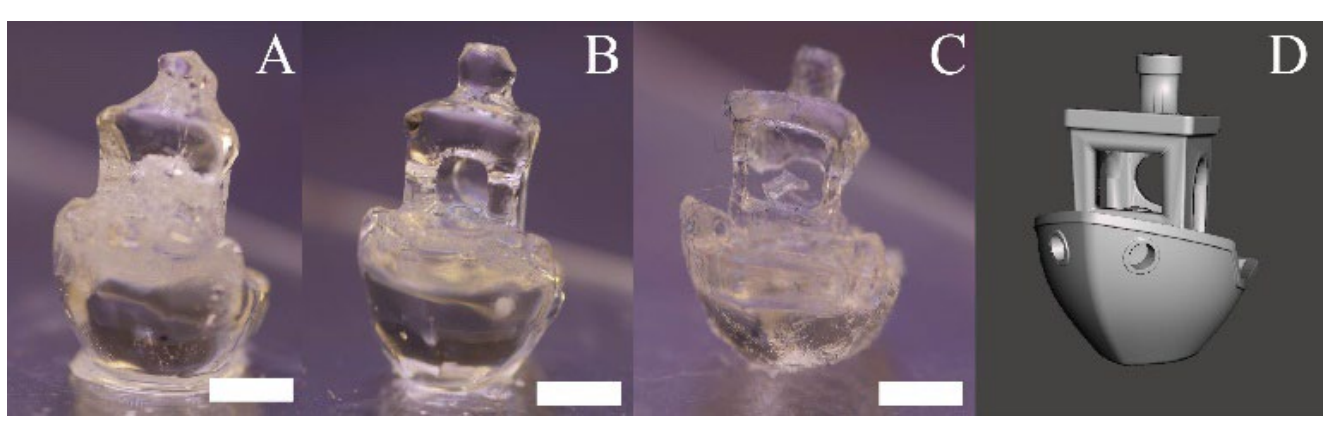

*Figure 6: Parts manufactured on laboratory setup post-processed with various liquid based cleaning methods (A) 30 minutes water with ultrasonic agitation. (B) 30 minutes water with 12.5% ethanol with ultrasonic agitation. (C) 30 minutes IPA mixing. (D) Reference geometry. Scalebars are 3 mm.*

However, liquid-based post-processing techniques inherently generate significant quantities of hazardous waste and do not readily allow for the reclamation of used material. To address these limitations, a second post-processing method was investigated based on the use of centrifugal forces to remove excess resin from both the vial and the surface of the printed part. In this method, the vial was oriented horizontally relative to gravity and rotated orthogonally about its axis, forcing the resin to exit through either of the vial's openings. Mesh barriers were installed at the vial ends to prevent the part from leaving the container during spinning. The expelled resin was collected at the bottom of the surrounding enclosure, enabling its recovery and reuse, as seen in Figure 7. Both of these post processing mechanisms proved successful in cleaning parts.

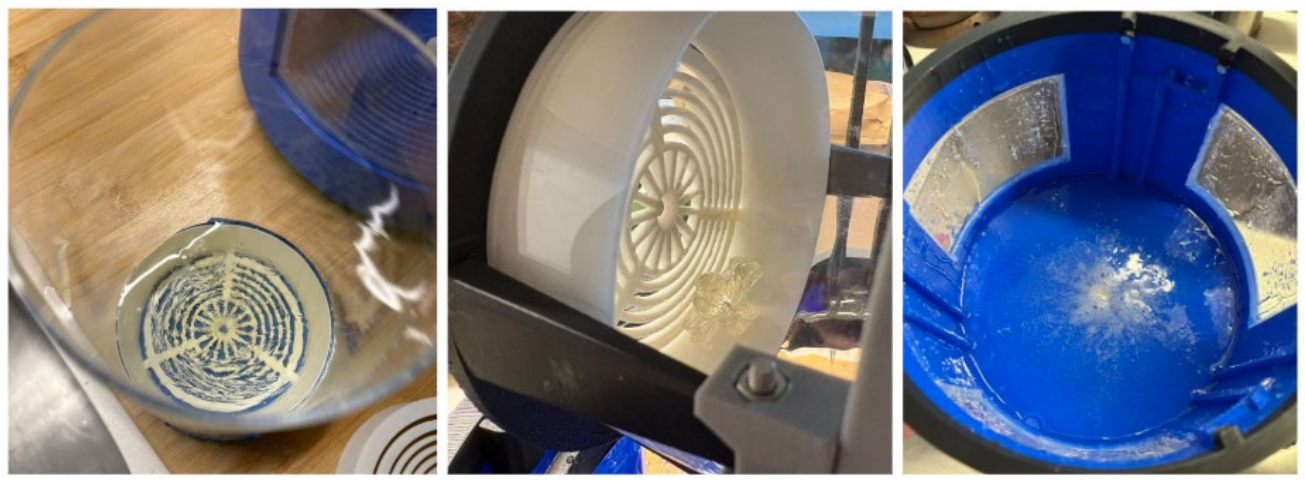
*Figure 7: The centrifugal system directly after a cleaning.*

### 2.5 Material Development

A major barrier to the broader adoption of CAL is the lack of accessible photopolymer materials. Currently, CAL-compatible materials are formulated within research laboratories, often using high-cost research-grade or custom-synthesized chemicals which limit the technology's scalability. As a first step toward mitigating this issue, several COTS prototyping resins were evaluated, including SLA-specific materials such as Formlabs Super Clear [11] and crafting resins such as Nicpro UV Crafting Resin [12]. Transmittance spectra were measured using a UV-Vis spectrophotometer across the 350–700 nm wavelength range. Nearly all COTS resins [12], [13], [14], [15], [16] exhibited near 100% transmittance at the projector's peak emission wavelength of 450 nm, as shown in Figure 8.

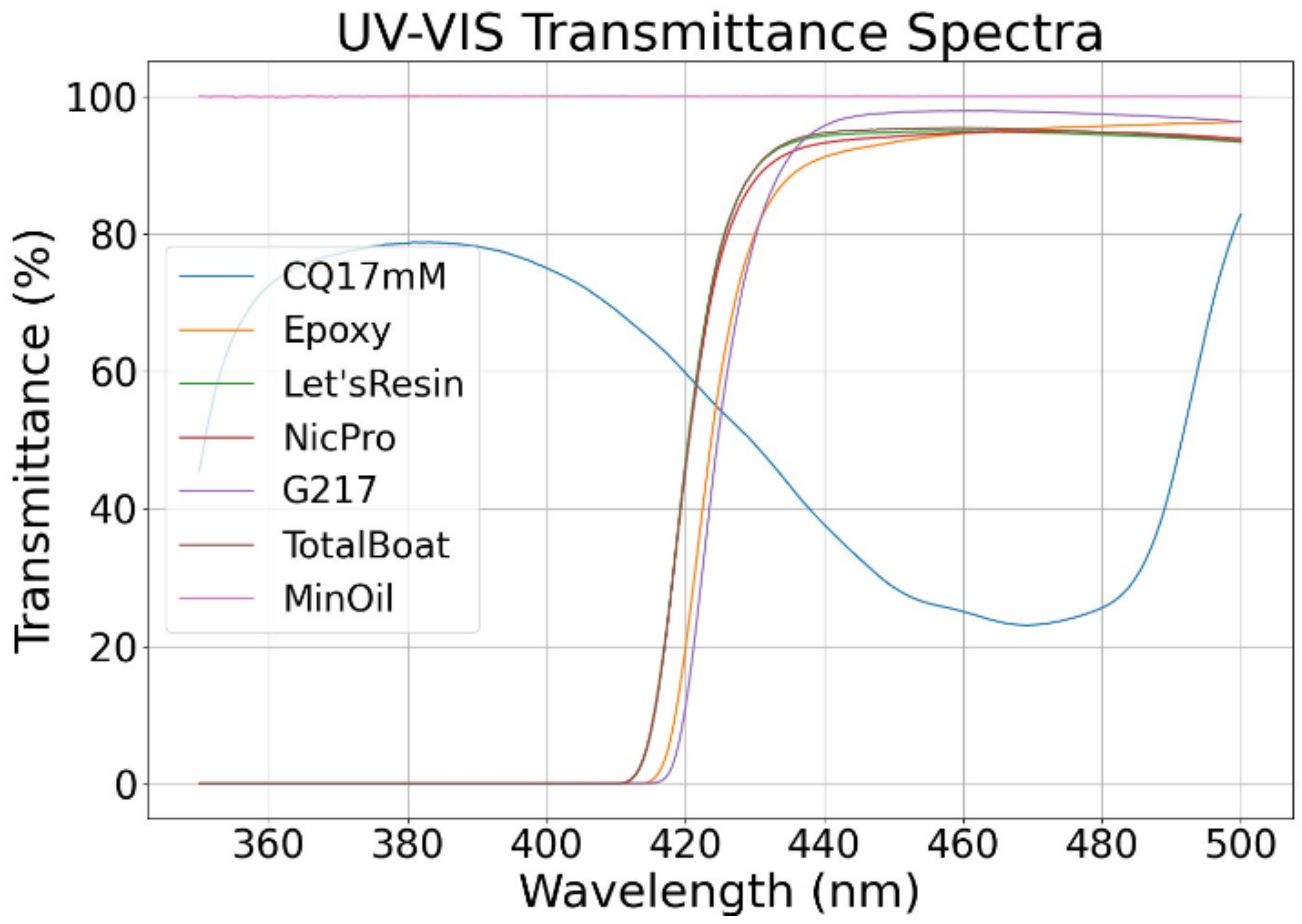

*Figure 8: Transmittance spectrum of various COTS resins and custom CAL CQ 17 mM resin.*

Given the insufficient absorption characteristics of existing COTS resins, a custom resin formulation was developed to maximize light absorption at the target wavelength. The resin was composed of a urethane dimethacrylate (UDMA) base (9000 cps viscosity) with 17 mM camphorquinone (CQ) as the photoinitiator and 8.5 mM ethyl 4-(dimethylamino)benzoate (EDAB) as the co-initiator. To support broad accessibility, efforts are underway in collaboration with a local resin supplier to produce and distribute the custom formulation as a commercially available COTS resin for the OpenCAL platform, with an expected cost of less than $100 per liter. Additionally, formulation and storage instructions will be provided for users who are unable to purchase the resin or who have the means to mix it independently. Fabrication requires only basic chemistry equipment commonly found in a high school laboratory—such as a hot plate, beakers, scale, and fume hood—though the initial investment in these materials may be more costly.

Initial validation confirmed the resin's compatibility with OpenCAL, as demonstrated by the successful fabrication of a complex geometry, shown in Figure 9.

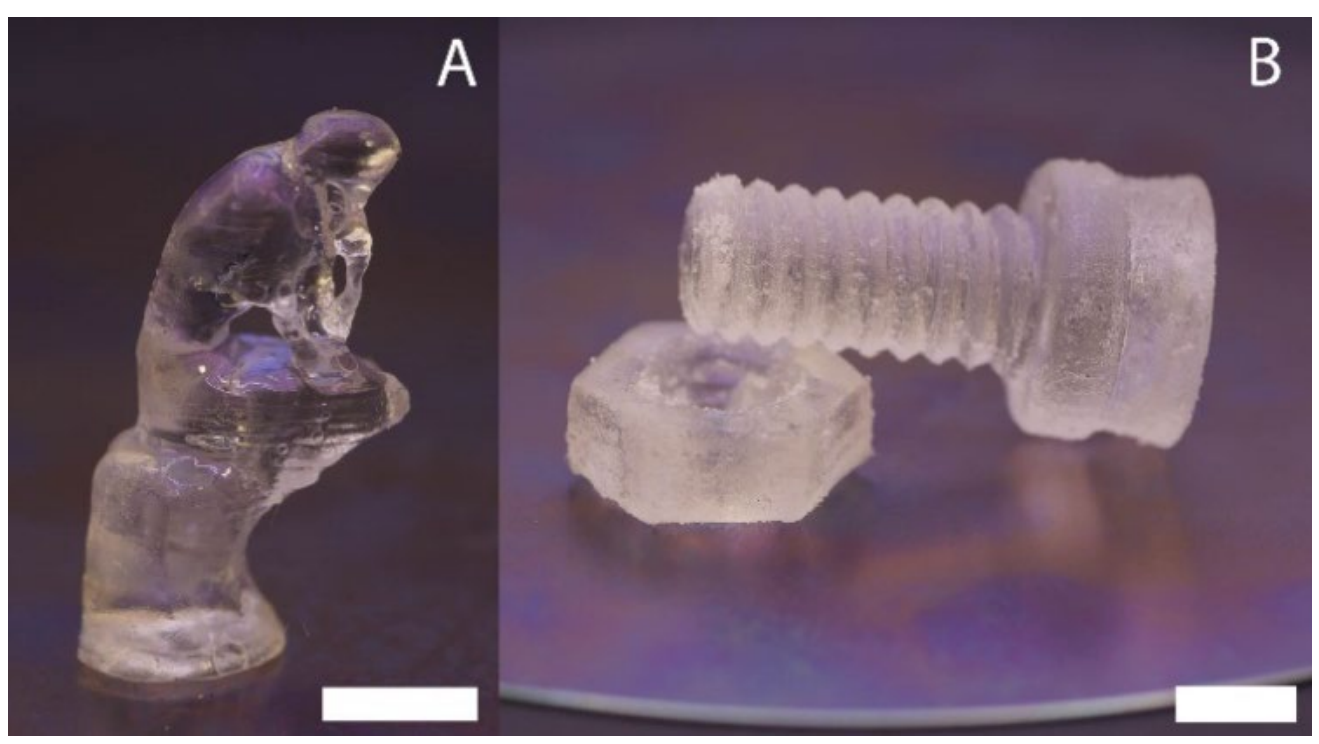

*Figure 9: Two geometries manufactured on the OpenCAL system with a 30 mm diameter vial. (A) The Thinker in UDMA. (B) A nut and bolt in UDMA. Scale bars are 5 mm.*

### 2.6 Community Development

The final core component of OpenCAL development is ensuring its successful adoption across multiple user communities. To facilitate this, a dedicated Discord server was established to provide a platform where interested users can ask questions, share feedback, and receive technical support. Access information is available upon request from the authors. The community has already contributed to the advancement of OpenCAL, including assisting with the development of independent systems, such as one built by two high school students in the Czech Republic. A major focus within the community has been the navigation and use of VAMToolbox, which is currently undergoing revisions to improve accessibility and user experience.

In addition to community support, multiple publicity avenues are being pursued to promote OpenCAL. A partnership with Make [17] is underway to publish a detailed guide on assembling and operating the system. Furthermore, a collaborative video with the 3D Printing Nerd YouTube channel [18] is planned to provide visual assembly instructions and offer additional user support. OpenCAL will also be directly demonstrated at major maker community events, including OpenSauce [19] and the Bay Area Maker Faire [20], to further engage and expand the user base. For academic and research groups, events such as the VAM Symposium serve as key venues for gathering feedback on OpenCAL and getting support from members of the high-technical community.

Finally, a pilot program at UC Berkeley's Jacobs Makerspace with UC Berkeley's Jacobs Makerspace [21] will evaluate OpenCAL's functionality in an academic setting. In this program, students will operate and modify the OpenCAL system under supervision, to analyze the technology's ability to function in an academic makerspace.

### Results & Discussion

The OpenCAL project demonstrated significant success in technological development and accessibility. Using only COTS components and standard makerspace equipment, a user-friendly CAL system was assembled with a final cost under $800 USD (Figure 10). The system fabricated the complex geometry shown in Figure 9 within 90 seconds, validating its optical and mechatronic accuracy. Nonetheless, further optical upgrades are required to fabricate larger parts, as the current beam divergence remains too high. Post-processing with the ethanol-based sonication method was completed within 20 minutes, effectively preserving fine features. OpenCAL met nearly all initial user-derived requirements. This performance demonstrates that OpenCAL effectively translates high-end volumetric additive manufacturing techniques into an accessible, low-cost platform suitable for academic environments.

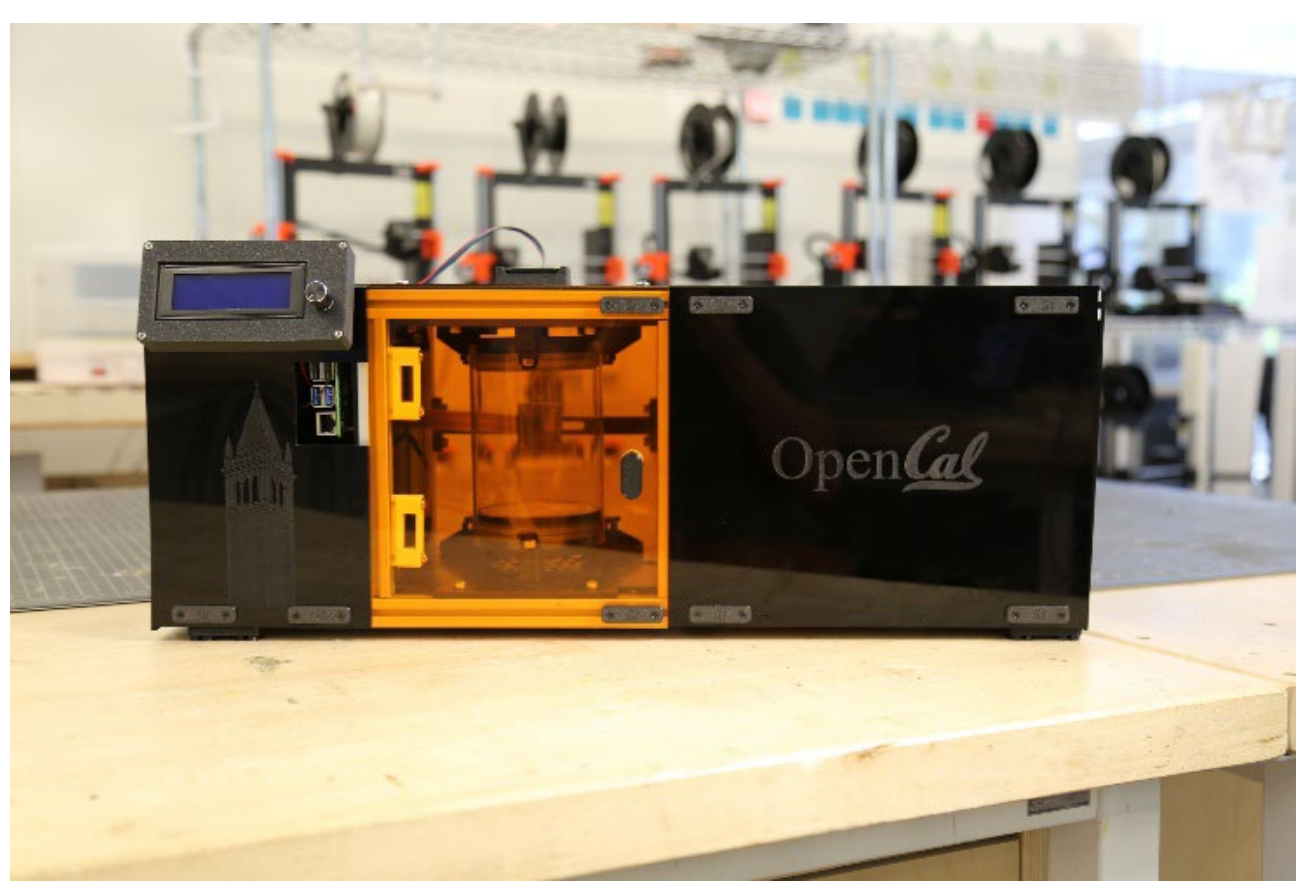

*Figure 10: OpenCAL Version One.*

A major barrier—availability of VAM materials—was addressed through the development of a standardized resin now commercially available. While initial mechanical characterization of printed parts was not conducted in this study, future work will include quantifying mechanical properties to ensure suitability for functional prototyping.

Community engagement has also been strong, with over 150 members on the OpenCAL Discord server and at least two groups developing independent CAL systems using COTS components. OnShape [22] will be used to share Computer Aided Design (CAD) documents, with a bill of materials, assembly and operational instructions being made into a ReadtheDocs [23] website. Both of these platforms encourage community contributions and edits.

The system was developed entirely by students, most of whom had no prior CAL experience, demonstrating that student-led teams can create impactful research platforms. By project completion, students were proficient in system operation and had proposed numerous design improvements. In addition to product development, they gained interdisciplinary skills in optical engineering, mechatronics, and photopolymer chemistry.

Challenges remain, particularly improving the usability of VAMToolbox software, which is critical for generating tomographic projection files. Future development priorities identified by the student team include automated print stopping, material handling, and enhanced user safety features. Curriculum modules are under development to support both system replication and understanding of core technical concepts such as optical system design. These initiatives aim to enhance the accessibility and educational value of OpenCAL, further strengthening its role as an interdisciplinary platform for collaborative learning and research development.

Because OpenCAL is composed of COTS components and uses accessible design practices, ongoing maintenance and upgrades can be performed by successive cohorts of students, promoting a self-sustaining educational ecosystem within makerspaces.

## Conclusion

The OpenCAL project successfully developed a low-cost, accessible volumetric additive manufacturing platform, meeting key technical requirements and addressing material accessibility barriers. By leveraging standard makerspace tools and COTS equipment, OpenCAL enables academic makerspaces to integrate advanced fabrication technologies previously confined to specialized research laboratories. The system also empowers makerspaces to contribute directly to the advancement of volumetric additive manufacturing by enabling students to participate in the open-source community, providing upgrades and innovations to the platform. OpenCAL fosters interdisciplinary education, offering students hands-on experience in optics, materials science, and computational imaging. Pilot deployments at academic institutions and other organizations will evaluate system robustness and educational impact. Pending successful outcomes, a broader public release is planned for late Q3 2025 to expand OpenCAL access across academic and maker communities.